# Quantum Key Recycling can share key more efficient than BB84

Yu-Chin Lu[1][0000-0001-5982-5611], Chia-Wei Tsai[0000-0002-8075-6558], and

Tzonelih Hwang[2][0000-0003-3881-1362]

[1,3]Department of Computer Science and Information Engineering, National Cheng Kung University, No. 1, University Rd., Tainan City, 70101, Taiwan, R.O.C.
[2]Department of Computer Science and Information Engineering, National Taitung University, No. 369, Sec. 2, University Rd., Taitung City, 95092, Taiwan, R.O.C.
hwangtl@csie.ncku.edu.tw

**Abstract.** We calculate the key sharing rate of Lu et al.'s Quantum Key Recycling (QKR) protocol. The key sharing rate is another version of the key rate, but it can be calculated for both the Quantum Key Distribution (QKD) protocols and the QKR protocols. We define the key sharing rate in this study and compare the key sharing rate of the QKR protocol to the rate of the QKD protocols. We found Lu et al.'s QKR protocol can be used to share keys more efficiently than BB84 in some situations. We also compare the six-state version of Lu et al.'s QKR protocol to the six-state QKD protocol. The results of this study show the potential advantages of using pre-shared keys to replace the public discussion in quantum protocols.

## 1 Introduction

### 1.1 Background

An important application of quantum cryptography is key distribution. A class of quantum protocol called Quantum Key Distribution (QKD) is designed to do this. The goal of QKD is to let all participants share secret keys. In 1984, the first QKD protocol BB84 [1] was proposed. After this, many studies proposed different QKD protocols [2-6]. An important research direction of QKD is to increase its efficiency with a noise channel. In BB84, the sender, Alice, randomly sends qubits in two bases (four state) $\{|0\rangle, |1\rangle, |+\rangle = \frac{1}{\sqrt{2}}(|0\rangle + |1\rangle), |-\rangle = \frac{1}{\sqrt{2}}(|0\rangle - |1\rangle)\}$ to the receiver, Bob. Bob will randomly choose a basis to measure the qubits. Then Alice and Bob will use a classical authentication channel to do the public discussion. If Alice and Bob choose the same basis to prepare/measure a qubit, the measurement result of this qubit will be their shared raw key. Otherwise, the measurement result will be discarded. When the quantum channel has noise, the qubits may be disturbed by noise and cause some error in the raw key. The error can also be caused by the eavesdropper, Eve, who wants to eavesdrop the communication. So Alice and Bob will do error correction and then privacy amplification to extract the secret key from the raw key. The ratio of the extracted secret key and the raw key is called key rate. The key rate is according to the error rate of the quantum channel. the key rate is an important efficiency index of a QKD protocol.

On the other hand, a class of quantum protocol called Quantum Key Recycling (QKR) is first proposed in 1982 [7]. The QKR protocol is designed to transmit a secret message. In the QKR protocol, the key used to encrypt the message can be recycled. Since the random number (which can be a key) is also a kind of message, we can use the QKR protocol to do the key distribution like the QKD protocol. The number of studies of the QKR protocol is fewer than the studies of the QKD protocol, and there is no QKR protocol that can run in the noise quantum channel[1]. Until 2020, Lu et al. propose a QKR protocol [12], which can run in the noise quantum channel and recycle the used key according to the error rate of the quantum channel.

In Lu et al.'s QKR protocol, Alice and Bob pre-shared three keys: $u$, $K_b$, $K_v$. Alice encodes the cipher $C = K_v \oplus ECC_d(m||MAC)$, where $m$ is the message, $MAC$ is the Message Authentication Code of $m$ and $ECC_d()$ is an error correction code encoding function, into two bases single qubits according to the key $K_b$ and sends the qubits to Bob. Bob measures the received qubits according to the key $K_b$, decoding the error correction code and then check the $MAC$. Alice and Bob then will recycle $u$, $K_b$ and partial recycle $K_v$ according to the error rate of the quantum channel. The ratio of the recycled key and the used pre-shared key is called key recycling rate.

Because Lu et al.'s QKR protocol can be used to do the key distribution (key sharing) with a noisy quantum channel, we can compare the efficiency of QKD protocols with the efficiency of Lu et al.'s QKR protocol.

### 1.2 Conclusion

In this study, we give a new definition of the efficiency of the QKD/QKR protocols called key sharing rate. And we compare the key sharing rate of Lu et al.'s QKR protocol with the key sharing rate of BB84. We find out that Lu et al.'s QKR protocol has a higher key sharing rate than BB84 when the error rate of the quantum channel is fixed and less than 0.11. We also modify Lu et al.'s QKR protocol to the six-state version of Lu et al.'s QKR protocol, which using three bases (sit-state) quantum state to encode the cipher and compare it with six-state QKD protocol [2, 3]. We find out that the QKR protocol gets more benefit from the six-state encoding than the QKD protocol. These results show that using the QKR protocol to share keys is a viable option.

---

[1] There are some studies proposed QKR protocols [8-11] which can run in the noise quantum channel. But they all have a common security loophole.

## 2 Preliminaries

### 2.1 Error Correction Code

The Error Correction Code (ECC) can be used to detect and correct errors of the message. ECC uses redundant information to correct errors. In this study, we use binary linear codes to be the ECC. We define an encoding function $ECC_d(M) = C$. This function can encode a message $M$ into a binary linear code $C$ with minimum Hamming distance $d$. We also define a decoding function which can decode the message $M$ from $C$, and detect and correct the potential errors. A binary linear code $C$ with minimum Hamming distance $d$ can correct at most $\lfloor \frac{d-1}{2} \rfloor$ errors. We also assume $C = M||s$, where $s$ is the redundant information added by $ECC_d()$ called syndrome.

If we want to correct more errors, we need to encode the message $M$ into a code $C$ with bigger $d$. The bigger the Hamming distance $d$, the longer the length of the code $C$. We can use information theory to calculate how many redundant information we need. When the error rate is $Q$, the redundant information needed to correct these errors is at least $H(Q)$, where $H()$ is the Shannon entropy in bits. The length of the redundant information in bits is also $H(Q)$ bits.

### 2.2 Key rate

In the QKD protocols, all participants will share m-bits raw key (a binary random number which may have errors and be partially knew by Eve) after the quantum part. The participants will do error correction and privacy amplification to extract a shorter but secure secret key, which is n-bits long. The key rate of the QKD protocols can be defined as $\frac{n}{m}$.

The meaning of the key rate is how many percents of the raw key can be used to be a secret key. The key rate is dependent on the error rate of the quantum channel in this round of the QKD protocol. The errors may come from the environment noise of the quantum channel or Eve's attack, and the participant can use public discussion to get the error rate. Since every error may come from Eve's attack, we need to consider all errors is caused by Eve to cover the worst case. So the meaning of the key rate is also how much information about the raw key is known by Eve.

The key rate is a useful efficiency index of the QKD protocol. At a certain error rate, a QKD protocol which has higher key rate can share more secret key and is more efficient. The error rate where the key rate goes to zero is also important. It shows the maximum error rate of the quantum channel that the QKD protocol can still share secret keys, for example, BB84 can share keys if the error rate is not beyond 0.11 and sit-state protocol can share keys if the error rate is not beyond 0.126.

### 2.3 Key Recycling Rate

The key recycling rate of Lu et al.'s QKR protocol is defined as below: Alice and Bob use $m$-bits pre-shared key to run the QKR protocol. After the protocol, they will use privacy amplification to shoring the used key to ensure the security of the key. The shorted key is $n$-bits, and the key recycling rate is $\frac{n}{m}$.

The key recycling rate looks like the key rate. They both use privacy amplification to ensure the security of the key. But the key rate is for the new shared key, and the recycling rate is for the used pre-shared key. Because of this difference, they cannot be compared directly.

## 3 Compare BB84 with Lu et al.'s QKR protocol [12]

Since QKR protocols are different from QKD protocols and can't calculate the "key rate" of a QKR protocol with the original definition of the key rate, we need a new definition of key rate to let us compare QKR protocols and QKD protocols.

**Definition 1.** *After the participants finish the quantum part of the QKR/QKD protocol, they will share an $m$-bits raw key. The participant will do the error correction and privacy amplification to extract a $n$-bits secret key, which is secure and synchronous between all participants. And these processings may consume $k$-bits pre-shared key. The key sharing rate is defined as $\frac{n-k}{m}$.*

For QKD protocols, the key sharing rate is just equal to the key rate. Since the QKD protocol doesn't need any pre-shared keys, it will consume no keys. The key sharing rate of a QKD protocol can be denoted by: $\frac{n}{m}$, which is the definition of the key rate of the QKD protocol.

For QKR protocols, they don't need to do privacy amplification to shorting the raw key, but they need to consume some keys to share the raw key. The key sharing rate of a QKR protocol is equal to $\frac{m-k}{m} = 1 - \frac{k}{m}$.

We then can compare the key sharing rate of BB84 with the key sharing rate of Lu et al.'s QKR protocol. The key sharing rate of BB84 just equals the key rate of BB84, and it is calculated by many studies [13, 14]. So we just need to calculate the key sharing rate of Lu et al.'s QKR protocol.

To calculate the key sharing rate of a QKR protocol, we need to know how many pre-shared key $k$ is consumed when Alice and Bob share $m$-bits raw key at quantum bit error rate (QBER) $Q$. In Lu et al.'s study, they already calculate the key recycling rate of their protocol[2] (as shown in Fig. 1). But the length of the consumed pre-shared key is according to not only the key recycling rate but also the length of used keys. In Lu et al.'s QKR protocol, a message $M$ contains the message ($m$-bits random number), $MAC$, and ECC. When the message is long enough, we can ignore the length of $MAC$. The length of ECC is according to the predicted noise level of the quantum channel $Q_{predict}$. Because the ECC should correct itself, the length of the ECC is $\frac{H(Q_{predict})}{1-H(Q_{predict})} * m$. Combine the above. We can get the length of $M$, and thus the length of the used key is $(1 + \frac{H(Q_{predict})}{1-H(Q_{predict})}) * m$. The length of the consumed key is then

$$\left(1 + \frac{H(Q_{predict})}{1-H(Q_{predict})}\right) * m * (1 - key\ recycling\ rate). \tag{1}$$

---

[2] In Lu et al.'s QKR protocol, there are three keys $u, K_b, K_v$ will be consumed. But when the message is long enough, the loss of the keys $u, K_b$ is less and can be ignored. In this study, we set the key recycling rate of $K_v$ as the key recycling rate of Lu et al.'s QKR protocol.

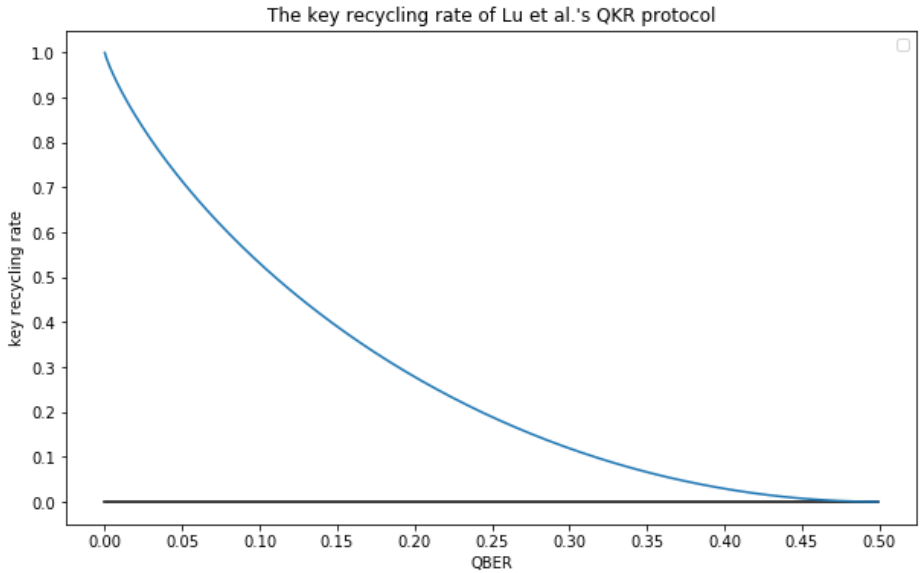

**Fig. 1.** The x-axis denotes the real QBER, and the y-axis denotes the key recycling rate of Lu et al.'s QKR protocol.

Now we are ready to calculate the key sharing rate of Lu et al.'s QKR protocol. Unlike the QKD protocol that estimates the QBER after the quantum part of the protocol by the public discussion, Lu et al.'s QKR protocol needs to predict the QBER before the quantum part of the protocol and decides how many error-tolerable we need. We can see Eq. (1) has two variable: $Q_{predict}$ and $Q$ (in the formula of key recycling rate). If the prediction QBER $Q_{predict}$ is less than the real one $Q$, Lu et al.'s QKR protocol will fail, and the key recycling rate is 0. On the other hand, if $Q_{predict}$ is more than $Q$, we have to waste some keys to protect the useless part of the syndrome and thus decrease the key sharing rate. When the prediction is perfect, we can get the best key sharing rate at that QBER. The result of the calculation of the key sharing rate of Lu et al.'s QKR protocol shows in Fig 2.

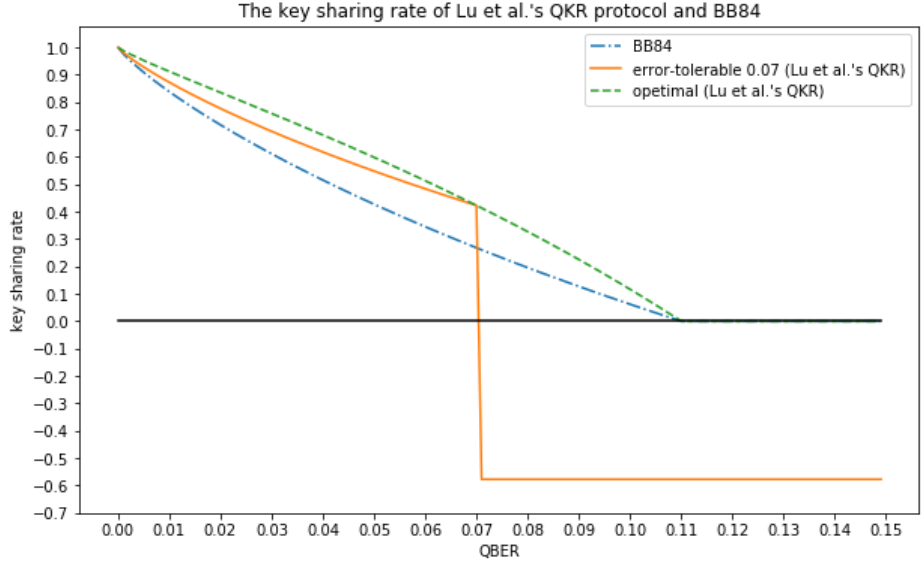

**Fig. 2.** The x-axis denotes the real QBER $Q$, and the y-axis denotes the key sharing rate. The line of error-tolerable 0.07 shows the key sharing rate of Lu et al.'s QKR protocol with the prediction QBER $Q_{predict} = 0.07$ according to $Q$. When $Q = 0.07$, this is the best key sharing rate out of all $Q_{predict}$ for Lu et al.'s QKR protocol. The line of optimal shows the best key sharing rate of Lu et al.'s QKR protocol we can achieve according to $Q$ (i.e., the key sharing rate when $Q_{predict} = Q$). And the line BB84 shows the key sharing rate (key rate) of the BB84 protocol according to the real $Q$.

In Lu et al.'s QKR protocol, the key $K_v$ used to protect the syndrome can partially recycle and thus is more efficient than the BB84 protocol, which leaks all information contained in the syndrome. But the length of the syndrome used in Lu et al.'s QKR protocol is longer than the syndrome used in the BB84 protocol because the BB84 protocol uses a classical authentication channel to send the syndrome and thus the syndrome doesn't need to correct itself. We can see when the QBER is low, the effect of the longer syndrome is not obvious, and Lu et al.'s QKR protocol has a higher key sharing rate. The high key sharing rate in Lu et al.'s QKR protocol is achieved at the cost of low error-tolerable and high cost of the keys derived from situations where the real QBER is higher than the prediction. It worth mentioning that when the prediction QBER is 0.11, the key sharing rate of our QKR protocol is just equal to the key sharing rate of BB84 and thus has no benefit. This shows that Lu et al.'s QKR protocol is a viable option. When the noise of the quantum channel is fixed and lower than 0.11, using Lu et al.'s QKR protocol to share keys is more efficient.

## 4   Variance

In this section, we modify Lu et al.'s QKR protocol to try to increase its key sharing rate. Unlike the original version of Lu et al.'s QKR protocol encoding the message into two bases quantum state, we encode the message into three bases (six-state) $\{|0\rangle, |1\rangle, |+\rangle, |-\rangle, |+y\rangle = \frac{1}{\sqrt{2}}(|0\rangle + i|1\rangle), |-y\rangle = \frac{1}{\sqrt{2}}(|0\rangle - i|1\rangle)\}$ in our modified QKR protocol, which we call six-state version of Lu et al.'s QKR protocol. This modified causes two effects of the key using in Lu et al.'s QKR protocol. First, we need more $K_b$ to decide the basis of every qubit. Second, the key recycling rate of $K_v$ is affected. The first difference does not influence the key recycling rate of $K_b$ in Lu et al.'s QKR protocol because the leakage of $K_b$ is only according to Bob's response. The effect of the second difference is shown in Fig. 3. The key recycling rate of $K_v$ is increased when the QBER does not achieve 0.5.

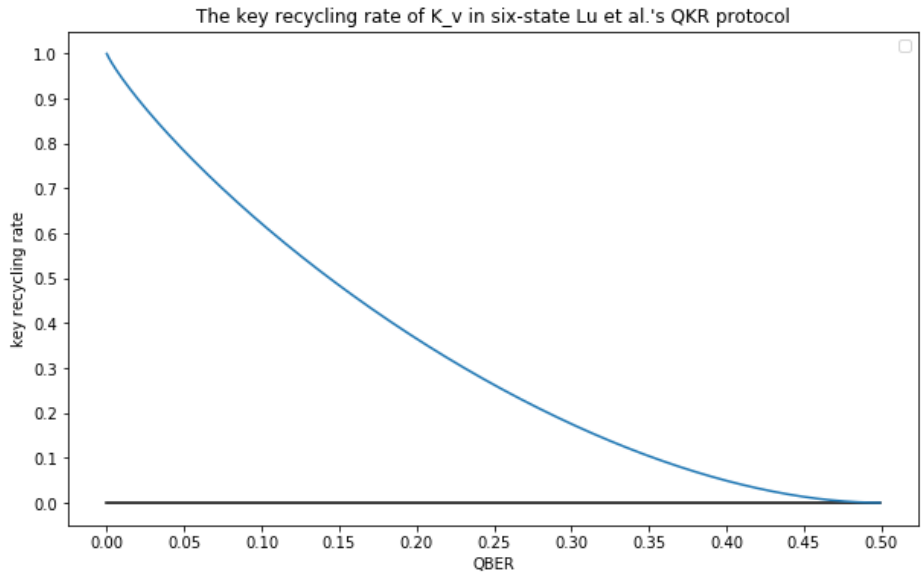

**Fig. 3.** The key recycling rate of $K_v$ in the six-state version of Lu et al.'s QKR protocol. Because the key recycling rate of $u$ and $K_b$ is almost 1, the key recycling rate of $K_v$ can also be seen as the total key recycling rate of the six-state version of Lu et al.'s QKR.

Because of the higher key recycling rate, the six-state version of Lu et al.'s QKR has a higher key sharing rate. We compare the sit-state version of Lu et al.'s QKR protocol with the six-state QKD protocol [2, 3]. The six-state QKD protocol is a modified version of BB84, which also using three bases (sit-state) qubits and has a higher key sharing rate (key rate) than BB84.

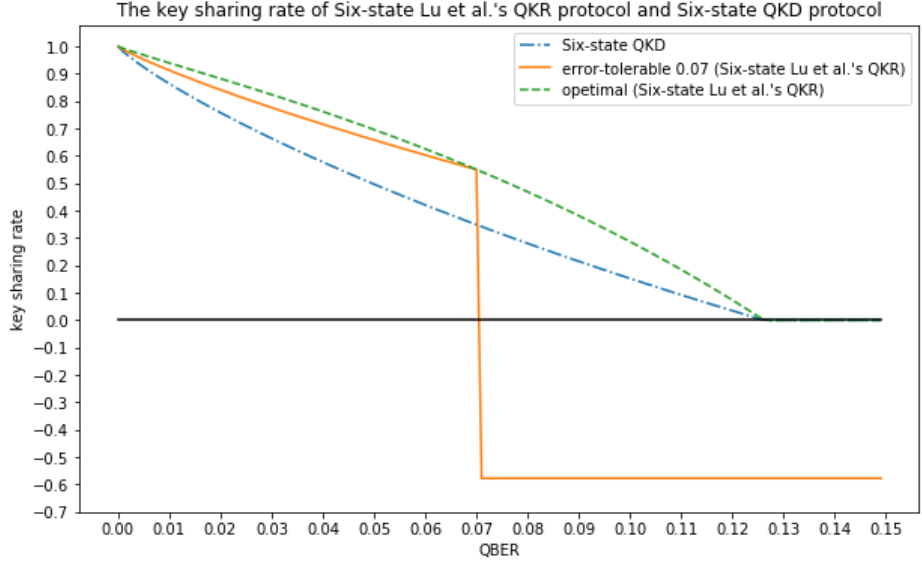

**Fig. 4.** The x-axis denotes the real QBER $Q$ and the y-axis denotes the key sharing rate. The line of error-tolerable 0.07 shows the key sharing rate of the six-state version of Lu et al.'s QKR protocol with the prediction QBER $Q_{predict} = 0.07$ according to $Q$. When $Q = 0.07$, this is the best key sharing rate out of all $Q_{predict}$ for six-state version of Lu et al.'s QKR protocol. The line of optimal shows the best key sharing rate of the six-state version of Lu et al.'s QKR protocol we can achieve according to $Q$ (i.e., the key sharing rate when $Q_{predict} = Q$). And the line six-state QKD shows the key sharing rate (key rate) of the six-state QKD protocol according to the real $Q$.

The comparing results of the key sharing rate of the six-state version of Lu et al.'s QKR protocol and the key sharing rate of six-state QKD protocol is very like the comparing results of the key sharing rate of the original version of Lu et al.'s QKR protocol and the key sharing rate of BB84. But if we calculate the difference of the key sharing rate of Lu et al.'s QKR protocol and the QKD protocol. We can find that the six-state version of Lu et al.'s QKR protocol gets more benefit of the key sharing rate from the six-state encoding than the six-state QKD protocol (as shown in Fig. 5).

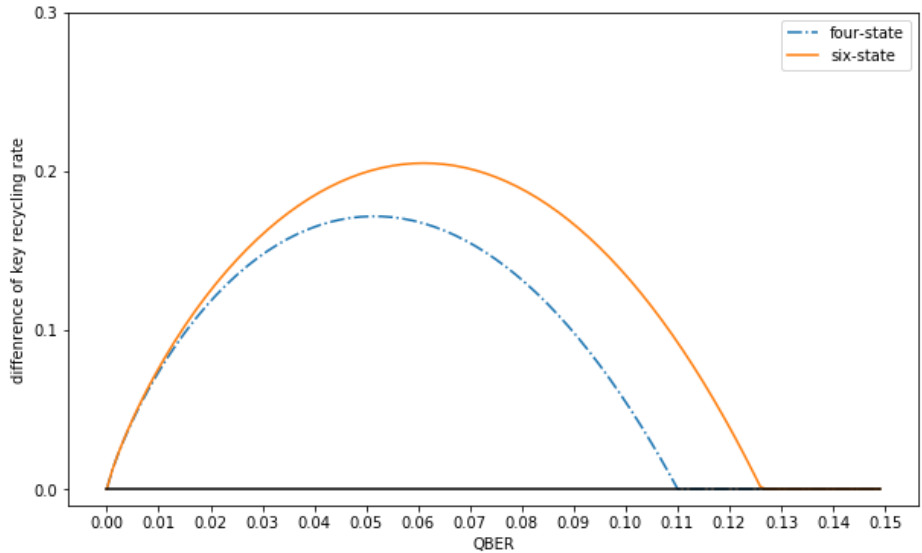

**Fig. 5.** The difference of the key sharing rate of Lu et al.'s QKR protocol and the QKD protocol. The line of four-state shows the difference in the key sharing rate of the original version of Lu et al.'s QKR protocol and BB84, according to the QBER $Q$. The line of six-state shows the difference of the key sharing rate of the six-state version of Lu et al.'s QKR protocol and six-state QKD protocol according to the QBER $Q$.

## 5   Conclusion and open questions

In this paper, we define the key sharing rate to compare the efficiency of key sharing of the QKD protocols and the QKR protocols. We calculate and compare the key sharing rate of Lu et al.'s QKR protocol, six-state version of Lu et al.'s QKR protocol, BB84, and the six-state protocol. We find out that when the noise level of the quantum channel is fixed and low, using Lu et al.'s QKR protocol to share key is more efficient than regular QKD protocols. This result shows the QKR protocol, which uses pre-shared keys to replace the public discussion, may get more advantage from the properties of quantum. And the benefit of the QKR protocol is not only to simplify the communication of the protocols [8, 15].

There is an open question of how to design a QKR protocol with a higher key sharing rate. In this study, we found out that the six-state version of Lu et al.'s QKR protocol has a higher key sharing rate than the original version. We think the way used to increase the key rate of the QKD protocol may also be used in the QKR protocol. This will be the research direction of our future work.